\documentclass{article}
\usepackage[T1]{fontenc}
\usepackage[utf8]{inputenc}
\usepackage{ismir} % Remove the "submission" option for camera-ready version
\usepackage{amsmath,cite,url}
\usepackage{graphicx}
\usepackage{color}
\usepackage{multirow}

\title{Enhancing Automatic Chord Recognition through LLM Chain-of-Thought Reasoning }

\multauthor
  {Chih-Cheng Chang$^1$ \hspace{1cm} Bo-Yu Chen$^2$ \hspace{1cm} Lu-Rong Chen$^2$ \hspace{1cm} Li Su$^2$}
  {
  $^1$ Institute of Information Science, Academia Sinica, Taiwan \\
  $^2$ Rhythm Culture Corporation, Taiwan\\
  {\tt\small ccchang12@iis.sinica.edu.tw}
  }

\def\authorname{C.-C Chang, B.-Y. Chen, L.-R. Chen and L. Su}

\usepackage[bookmarks=false,pdfauthor={\authorname},pdfsubject={\pdfsubject},hidelinks]{hyperref}
\begin{document}

\maketitle

\begin{abstract}
Music Information Retrieval (MIR) encompasses a broad range of computational techniques for analyzing and understanding musical content, with recent deep learning advances driving substantial improvements. Building upon these advances, this paper explores how large language models (LLMs) can serve as an integrative bridge to connect and integrate information from multiple MIR tools, with a focus on enhancing automatic chord recognition performance. We present a novel approach that positions text-based LLMs as intelligent coordinators that process and integrate outputs from diverse state-of-the-art MIR tools—including music source separation, key detection, chord recognition, and beat tracking. Our method converts audio-derived musical information into textual representations, enabling LLMs to perform reasoning and correction specifically for chord recognition tasks. We design a 5-stage chain-of-thought framework that allows GPT-4o to systematically analyze, compare, and refine chord recognition results by leveraging music-theoretical knowledge to integrate information across different MIR components. Experimental evaluation on three datasets demonstrates consistent improvements across multiple evaluation metrics, with overall accuracy gains of 1-2.77\% on the MIREX metric. Our findings demonstrate that LLMs can effectively function as integrative bridges in MIR pipelines, opening new directions for multi-tool coordination in music information retrieval tasks.\footnote{Code repository: \href{https://github.com/WildHoneyPie/ChordCoT}{https://github.com/WildHoneyPie/ChordCoT}}
\end{abstract}

\section{Introduction}\label{sec:introduction}

In tonal music systems, chords serve as the fundamental harmonic building blocks, providing the structural foundation that shapes musical expression and emotional content. The ability to identify and understand chord progressions plays a crucial role in music education, enabling students to develop deeper comprehension of harmonic relationships and musical structure. Given the importance of chords in musical understanding and practice, automatic chord recognition (ACR) has been a pivotal task in the field of music information retrieval (MIR). Recent advances in deep learning have revolutionized this field, leading to substantial improvements in recognition accuracy compared to traditional signal processing approaches \cite{chord}.

Parallel to these advances in audio-based deep learning, text-based Large Language Models (LLMs) pre-trained on massive text corpora have demonstrated remarkable proficiency in reasoning and understanding across various domains, including musical knowledge and comprehension. Recent works have shown the capability of zero-shot text-based LLM models to perform music understanding tasks without fine-tuning \cite{ChatMusician}. Furthermore, subsequent research has demonstrated the effectiveness of these LLM models in musical applications such as music generation \cite{LLM_musicgen}, audio effect prediction \cite{LLM2Fx}, and error detection in MIR tasks \cite{LLM_music_judge}. These findings suggest that LLMs' text-based reasoning abilities could enhance MIR task performance by processing music information that has been converted into textual representations.

This paper aims to investigate whether the capabilities of text-based LLMs can be leveraged to understand musical content and enhance the performance of the ACR task. We employ various state-of-the-art MIR tools, including music source separation \cite{demucs}, key detection \cite{keycnn}, chord recognition \cite{chord}, and beat tracking \cite{beat}, to extract musical information from audio signals. The outputs from these models are then converted into textual representations that are readable by text-based LLMs. Without training or fine-tuning any deep learning models, we utilize prompt engineering and chain-of-thought (CoT) reasoning techniques to enable language models to refine and improve chord recognition results \cite{cot}. This approach explores the potential of combining existing audio analysis tools with the reasoning capabilities of LLMs to achieve better performance in automatic chord recognition tasks. 

Figure 1 illustrates our proposed approach, which integrates outputs from multiple MIR tools through a systematic LLM-based reasoning framework. By converting audio-derived musical information into textual representations, our method enables GPT-4o to perform cross-modal analysis and refinement, demonstrating the potential of language models as intelligent coordinators in music information retrieval pipelines.

\begin{figure*}[!t]
  \centering
  \includegraphics[width=\textwidth]{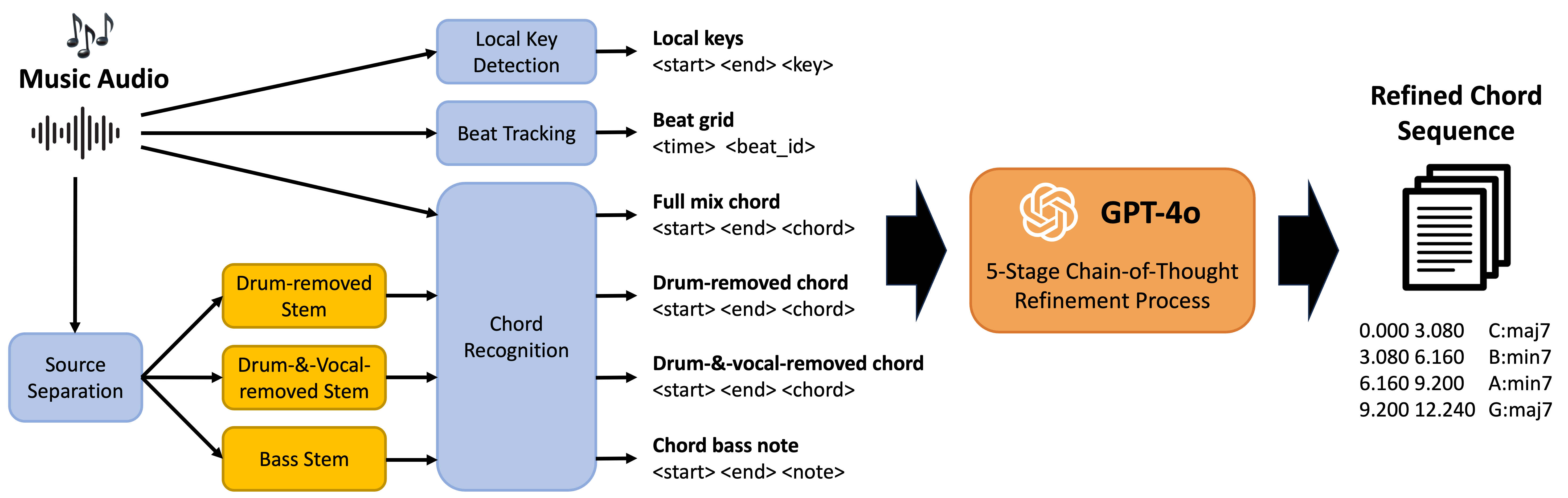}
  \caption{Overview of the proposed LLM-enhanced chord recognition system. The system converts outputs from multiple MIR tools (local key detection, beat tracking, source separation, and chord recognition) into standardized textual formats, which are then processed by GPT-4o through a 5-stage chain-of-thought refinement process to produce enhanced chord recognition results.}
  \label{fig:cot-pipeline}
\end{figure*}

\section{Methodology}\label{sec:method}
We employ GPT-4o \cite{gpt4} as an integrative bridge to connect and coordinate information from multiple MIR tools for enhanced automatic chord recognition. Our method converts audio-derived musical information into textual representations, enabling the LLM to perform reasoning and correction through a 5-stage chain-of-thought framework that systematically analyzes, compares, and refines chord recognition results.

\subsection{Audio Signal Analysis with MIR Tools}
To enhance automatic chord recognition performance, we employ multiple MIR tools to extract complementary musical information from audio signals. We first utilize HT Demucs \cite{demucs} for music source separation to generate separated audio tracks: drum-removed audio, drum-and-vocal-removed audio, and isolated bass stem. These separated tracks, along with the original full-mix audio, are then processed through  a large-vocabulary chord recognition model capable of recognizing 301 distinct chord classes \cite{chord}. This multi-track approach aims to obtain multiple chord recognition results, as different source combinations may yield varying accuracy depending on the musical content—the optimal configuration will be determined through systematic comparison in our refinement framework. For the bass stem specifically, we perform chord recognition but only retain the root/bass note information, discarding the chord quality components to avoid potential conflicts with the original chord sequence, as bass lines typically provide strong harmonic anchoring information that is most reliable for root note identification. To provide tonal context for music-theoretical reasoning, we employ local key detection \cite{keycnn} on the full-mix audio, which helps identify chord progressions that may be inconsistent with the detected key. For temporal alignment, we apply Beat This \cite{beat} to detect beat positions in the original audio, enabling precise synchronization of chord boundaries to musical beats. All outputs from these MIR tools are converted into standardized textual representations, allowing GPT-4o to integrate and reason about the multi-modal audio-derived information through our proposed 5-stage chain-of-thought refinement framework.

\subsection{5-Stage CoT Refinement}
\subsubsection{Stage 1: Music Source Separation (MSS)}
In this stage, we perform chord recognition on three audio tracks generated from the source separation process: the original full-mix audio, the drums-removed audio, and the drums-and-vocals-removed audio. Each track configuration offers different advantages—removing drums eliminates non-tonal percussive interference that may confuse the chord recognition model, while the drums-and-vocals-removed track focuses purely on instrumental harmony from bass and accompanying instruments. However, vocals can sometimes reinforce harmonic content through melodic lines and harmonies, making the full-mix or drums-removed versions potentially more informative in certain passages. The LLM processes the chord recognition results from all three audio tracks and determines which result is optimal based on three criteria: the proportion of 'N' (non-chord) labels, chord progression patterns, and music-theoretical plausibility. The selected best-performing result serves as the primary chord sequence, while the second-best result is retained as a reference for subsequent stages.

\subsubsection{Stage 2: Bass Correction}
In this stage, the LLM processes three inputs: the bass stem chord recognition results (with only root/bass note information retained), the local key estimation results, and the selected best-performing ACR results from Stage 1. Based on music theory knowledge, the LLM evaluates whether the bass stem chord recognition results are reliable. If deemed reliable, the process proceeds to the bass correction phase; otherwise, this stage is skipped. During the bass correction phase, the LLM refines the chord recognition results from the previous stage using the bass stem information, focusing specifically on bass note corrections. If the bass note corresponds to a chord tone of the original chord, the chord remains unchanged or undergoes appropriate inversion. If the bass note does not match the original chord tones but falls within the key, the LLM treats the bass note as the root and corrects the chord to the corresponding diatonic chord in the local key.

\subsubsection{Stage 3: Key Correction}
In this stage, the LLM evaluates the current chord recognition results based on local key estimation to identify potentially unreasonable segments. Additionally, the second-best chord recognition result from Stage 1 is used as a reference. When discrepancies exist between the current result and the reference, the LLM assesses whether the reference result is superior based on chord patterns and music theory principles. In the prompt design, we instruct the LLM to adopt a conservative correction approach, maintaining flexibility for modal interchange, secondary dominant chords, and other common harmonic practices. Corrections are applied only to segments that are clearly inconsistent with established music theory principles.

\subsubsection{Stage 4: Anomaly Detection}
Building upon the results from Stage 3, this stage instructs the LLM to identify unreasonable chord segments again based on local key estimation information and music theory knowledge. The LLM also evaluates whether certain segments labeled as 'N' (non-chord) should actually contain chord information. To enhance the LLM's reasoning process, we design a two-step approach on this stage: first, the LLM lists potentially problematic segments along with the reasons for identifying them as problematic; then, the LLM applies corrections based on the reasoning it has previously claimed. This explicit reasoning step improves the transparency, robustness, and accuracy of the correction process.

\subsubsection{Stage 5: Beat Alignment}
This stage performs beat alignment based on the beat tracking results. Unlike previous stages, this stage does not utilize LLMs but employs a rule-based approach for alignment. Each detected beat corresponds to a quarter note, which we subdivide into equal parts to calculate all sixteenth note positions. The start time and end time of each chord label from the final chord recognition results are directly aligned to the nearest sixteenth note beat positions. To prevent degradation from inaccurate beat tracking, we skip this stage if the required temporal displacement exceeds a predefined threshold.

\begin{table*}[ht]
\centering
\begin{tabular}{llccccccc}
\hline
\textbf{Dataset} & \textbf{Stage} & \textbf{MIREX} & \textbf{Root} & \textbf{Majmin} & \textbf{Thirds} & \textbf{Triads} & \textbf{Sevenths} & \textbf{Tetrads} \\
\hline
\multirow{6}{*}{IdolSongsJp \cite{jidol}}
  & Baseline \cite{chord}           & 79.50 & 80.91 & 80.52 & 77.73 & 74.93 & 65.72 & 59.19 \\
  & MSS                & 80.67 & 82.29 & 82.35 & 79.48 & 76.67 & 67.11 & 60.45 \\
  & Bass Correction    & 80.71 & 82.29 & 82.35 & 79.48 & 76.67 & 67.11 & 60.45 \\
  & Key Correction     & 80.69 & 82.29 & 82.35 & 79.48 & 76.67 & 67.11 & 60.45 \\
  & Anomaly Detection  & \textbf{81.16} & \textbf{82.77} & \textbf{82.82} & \textbf{79.92} & \textbf{77.11} & \textbf{67.48} & \textbf{60.80} \\
  & Beat Alignment     & 80.73 & 82.44 & 82.27 & 79.56 & 76.70 & 67.00 & 60.46 \\
\hline
\multirow{6}{*}{UsPop2002}
  & Baseline \cite{chord}            & 80.07 & 82.52 & 82.06 & 79.55 & 72.47 & 72.55 & 63.18 \\
  & MSS                & 80.85 & 83.22 & 82.94 & 80.04 & 73.16 & 73.20 & 63.69 \\
  & Bass Correction    & 80.84 & 83.22 & 82.92 & 80.04 & 73.13 & 73.06 & 63.54 \\
  & Key Correction     & 80.89 & 83.24 & 82.96 & 80.35 & 73.18 & 73.20 & 63.68 \\
  & Anomaly Detection  & 80.95 & 83.36 & 83.04 & 80.47 & 73.25 & 73.24 & 63.73 \\
  & Beat Alignment     & \textbf{81.13} & \textbf{83.54} & \textbf{83.21} & \textbf{80.64} & \textbf{73.42} & \textbf{73.39} & \textbf{63.87} \\
\hline
\multirow{6}{*}{In-house dataset}
  & Baseline \cite{chord}            & 83.29 & 80.25 & 80.79 & 79.48 & 78.89 & 67.17 & 64.52 \\
  & MSS                & 84.12 & 81.63 & 81.01 & 79.97 & 79.52 & 73.91 & 69.57 \\
  & Bass Correction    & 84.04 & 81.54 & 80.92 & 79.88 & 79.42 & 73.80 & 69.46 \\
  & Key Correction     & 85.29 & 82.29 & 82.28 & 81.13 & 80.67 & 74.15 & 69.50 \\
  & Anomaly Detection  & 85.93 & 83.05 & 82.63 & 81.57 & 80.98 & 74.49 & 69.81 \\
  & Beat Alignment     & \textbf{86.06} & \textbf{83.30} & \textbf{82.81} & \textbf{81.79} & \textbf{81.17} & \textbf{74.64} & \textbf{69.94} \\
\hline
\end{tabular}
\caption{Framewise chord recognition accuracy (\%) across different datasets and processing stages. "Original" represents the baseline ACR model results obtained directly from the original full mix music audio tracks. Bold values indicate the best accuracy for each metric within each dataset.}
\label{tab:chord_results}
\end{table*}

\section{Experiment}\label{sec:exp}

\subsection{Dataset}

Our experiments are evaluated on three datasets. The first dataset, IdolSongsJp, consists of 15 tracks created by commissioning professional composers to produce music in the style of Japanese idol groups \cite{jidol}. The second dataset comprises a subset of 192 full-length pop songs selected from UsPop2002\footnote{https://github.com/tmc323/Chord-Annotations}. The third dataset is an in-house collection containing 20 Western pop songs, featuring only chorus segments that have been professionally annotated with chord labels by expert musicians. These datasets are accompanied by label files that specify the start time, end time, and type of the chord.

\subsection{Baseline Chord Recognition Model}
We employ the large-vocabulary chord transcription system , which utilizes chord structure decomposition to handle extensive chord vocabularies as our baseline ACR model \cite{chord}. This baseline model is capable of recognizing 301 distinct chord classes, including the following categories: basic triads (maj, min, aug, dim), inverted triads (maj/3, maj/5, min/b3, min/5), seventh chords (maj7, 7, min7, dim7, hdim7), extended chords (maj9, 9, min9, 11, 13), suspended chords (sus4, sus2, sus4(b7)), and slash chords (maj/2, maj/b7, min/2, min/b7), along with a non-chord class (N). 

\subsection{Text-based Information Representation}
To enable LLM processing, all MIR tool outputs are converted into standardized textual formats using a consistent three-column structure: \texttt{<start\_time> <end\_time> <label>}. Chord recognition and key detection results follow the format with chord labels using the Harte shorthand notation system \cite{harte}. For example, "A:maj/3" represents an A major chord with the third in the bass, while "F:min7" denotes an F minor seventh chord. For bass stem analysis, we discard chord quality components to extract only the pitch calss.
Beat tracking results utilize the same temporal format with beat positions numbered 1-4, where "1" indicates the downbeat and "2", "3", "4" represent subsequent beats within each measure. Throughout all five stages, we maintain consistent adherence to these formats in both LLM inputs and outputs, ensuring seamless integration between processing stages.

\subsection{Evaluation Metrics}

For evaluation, we assess the performance of our proposed LLM-enhanced chord recognition system using framewise chord recognition accuracy. The metric calculates the percentage of correctly identified chord labels at each time frame throughout the audio tracks. All evaluation scores are computed using the \texttt{mir\_eval}  library \cite{mireval}, which provides standardized implementations of music information retrieval evaluation metrics. Specifically, the adopted metrics in this study are: root only, root and thirds, major/minor, triads, sevenths, tetrads and MIREX (considered correct if it shares at least three pitch classes in commons).

\subsection{Results and Discussion}
Table 1 shows the framewise chord recognition accuracy across different datasets and processing stages. Our proposed LLM-enhanced system demonstrates consistent improvements across various datasets and evaluation metrics. Most stages effectively enhance ACR accuracy, with the overall system achieving improvements of 1-2.77\% on the MIREX metric across all datasets, representing substantial gains in automatic chord recognition performance.

However, several noteworthy observations emerge from detailed analysis. The Bass Correction stage occasionally results in decreased accuracy across most scenarios. Upon examining individual songs, we find that while most tracks experience minor improvements or maintain stable performance, a subset of songs suffers from sustained bass notes that cause the system to incorrectly modify properly recognized chord changes. These corrections replace dynamic chord progressions with static diatonic chords corresponding to the persistent bass note. Since these erroneous corrections typically occupy longer durations, they exert a disproportionately dominant impact on overall dataset performance.

Our in-house dataset exhibits significantly more pronounced improvements compared to the other two datasets. We attribute this enhanced performance to two factors: the dataset contains only chorus segments, providing shorter contexts that enable more reliable LLM processing, and the increased pattern regularity in chorus sections allows the LLM to make more confident and accurate corrections.

During the beat alignment stage, IdolSongsJp represents the only dataset showing performance degradation. Analysis reveals that Japanese idol-style songs feature more frequent chord changes compared to conventional pop music, rendering beat alignment ineffective for this particular musical style with its characteristic rapid harmonic rhythm.

\section{Conclusion and Future Work}

This study presented a novel approach that leverages LLMs as integrative bridges to enhance automatic chord recognition by coordinating information from multiple MIR tools. Through our 5-stage CoT framework, a closed source GPT-4o model successfully demonstrated the ability to perform music-theoretical reasoning and systematically refine chord recognition results by integrating outputs from music source separation, key detection, chord recognition, and beat tracking systems. Experimental results across three diverse datasets show consistent improvements, with accuracy gains of 1-2.7\% on the MIREX metric, validating the effectiveness of using text-based reasoning to complement traditional deep learning approaches.

Building upon the success of this integrative approach, numerous avenues for enhancement and expansion present themselves. Incorporating music structure segmentation could enhance system stability by applying segment-specific refinement strategies, potentially making the approach more robust across different musical sections. The integration of music genre classification would enable genre-specific system prompts, allowing the LLM to apply more targeted musical knowledge and correction strategies tailored to different musical styles, from jazz harmonies to electronic music progressions.

Beyond chord recognition, this framework promises significant potential for broader cross-modal enhancement across other MIR tasks, including music transcription, music structure segmentation, and integrated multi-task MIR systems. This approach could advance toward more comprehensive language model integration in music information retrieval, where LLMs serve as central reasoning engines coordinating multiple specialized audio analysis tools.

% For BibTeX users:
\bibliography{ISMIR_LLM4Music_template}

% For non BibTeX users:
%\begin{thebibliography}{citations}
% \bibitem{Author:17}
% E.~Author and B.~Authour, ``The title of the conference paper,'' in {\em Proc.
% of the Int. Society for Music Information Retrieval Conf.}, (Suzhou, China),
% pp.~111--117, 2017.
%
% \bibitem{Someone:10}
% A.~Someone, B.~Someone, and C.~Someone, ``The title of the journal paper,''
%  {\em Journal of New Music Research}, vol.~A, pp.~111--222, September 2010.
%
% \bibitem{Person:20}
% O.~Person, {\em Title of the Book}.
% \newblock Montr\'{e}al, Canada: McGill-Queen's University Press, 2021.
%
% \bibitem{Person:09}
% F.~Person and S.~Person, ``Title of a chapter this book,'' in {\em A Book
% Containing Delightful Chapters} (A.~G. Editor, ed.), pp.~58--102, Tokyo,
% Japan: The Publisher, 2009.
%
%\end{thebibliography}

\end{document}